\documentclass{elsart}

\usepackage{times}
\usepackage{graphicx}
\usepackage{amssymb}

\begin{document}

\begin{frontmatter}

\title{A cluster algorithm for Lattice Gauge Theories}
\author[ia1,ia2] {Fabien Alet}
\ead{alet@phys.ethz.ch}
\author[ia1]{Biagio Lucini\corauthref{cor1}}
\corauth[cor1]{Corresponding Author}
\ead{lucini@phys.ethz.ch}
\author[ia1] {Michele Vettorazzo}
\ead{vettoraz@phys.ethz.ch}
\address[ia1]{Institut for Theoretical Physics, ETH Z\"urich,
CH-8093 Z\"urich, Switzerland}
\address[ia2]{Computational Laboratory, ETH Z\"urich,
CH-8092 Z\"urich, Switzerland}
\begin{abstract}
A new algorithm for simulating compact U(1) lattice gauge theory in
three dimensions is presented which is based on global changes in the
configuration space. We show that this algorithm provides an
effective way to extract partition functions at given external flux. As an
application, we study numerically the  finite temperature
deconfinement phase transition.
\end{abstract}
\end{frontmatter}
\begin{figure}[htb]
\begin{center}
\includegraphics[scale=0.4]{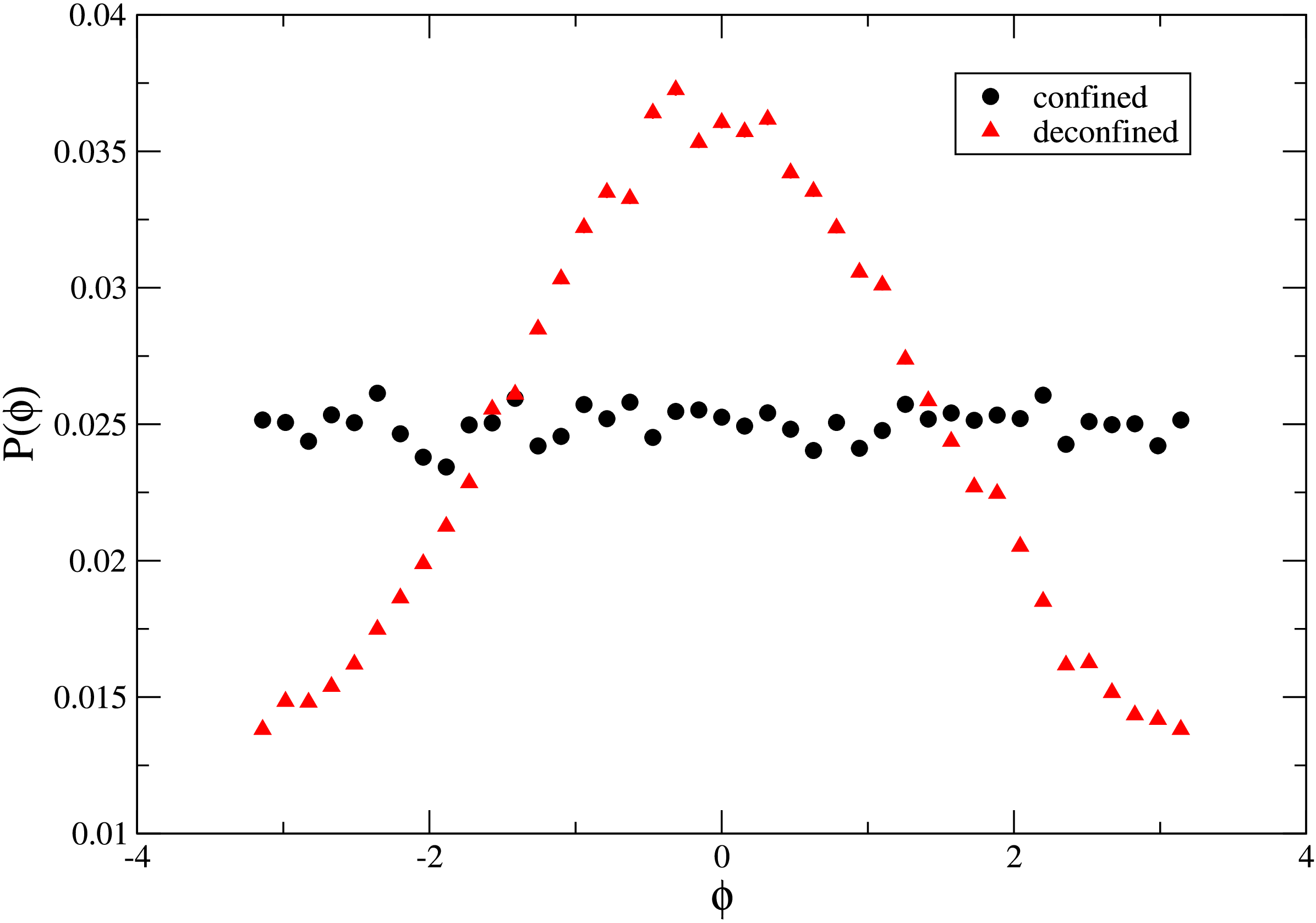}
\end{center} 
\caption{
Probability distribution of the flux on a $32^2 \times 6$ lattice
at $\beta = 1.80$ (confined phase) and $\beta = 2.50$ (deconfined phase).}
\label{fig1}
\end{figure}
\begin{figure}[htb]
\begin{center}
\includegraphics[scale=0.4]{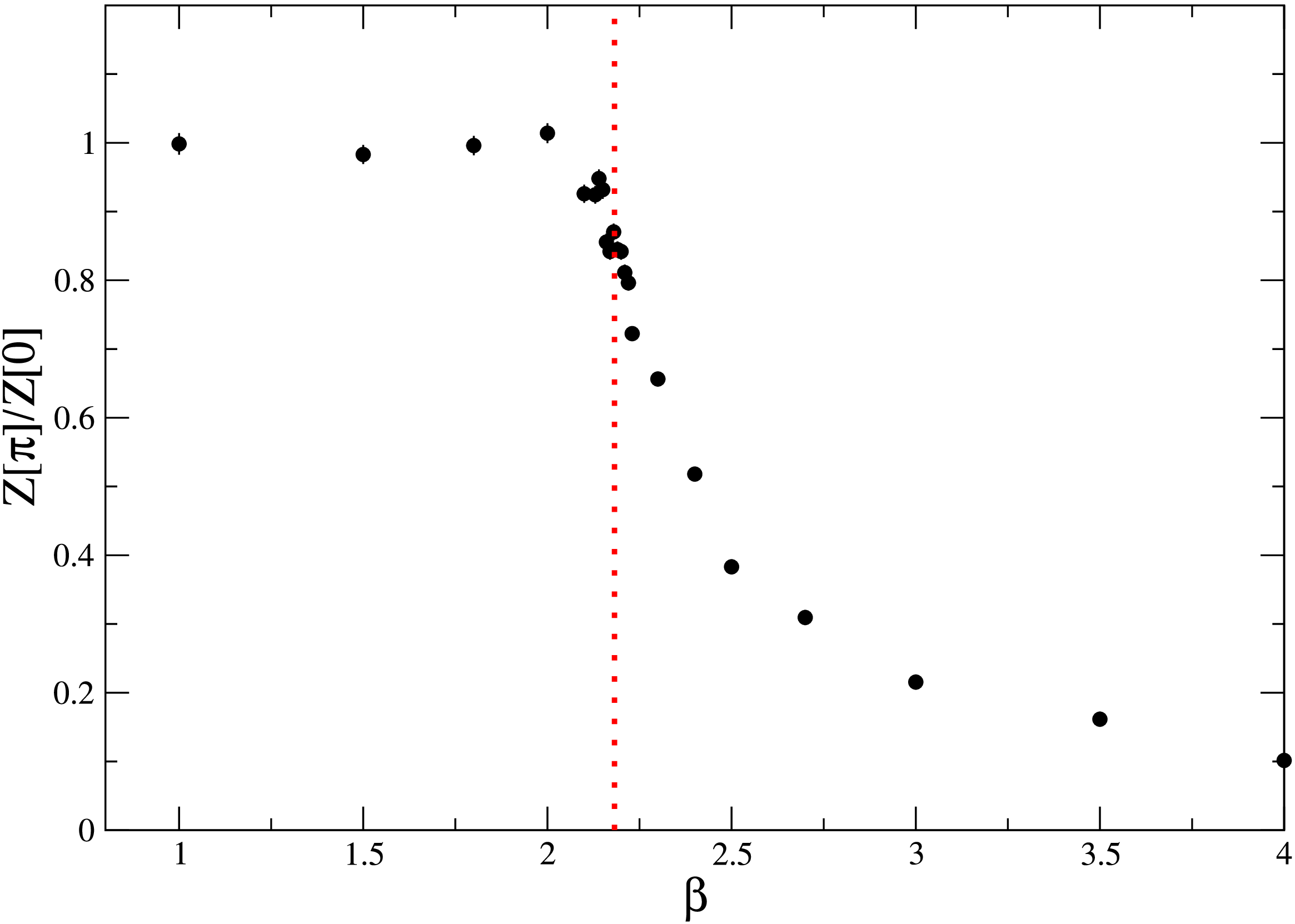}
\end{center} 
\caption{$Z[\pi]/Z[0]$ as a function of $\beta$ on a $32^2 \times 6$.
The vertical line is located at $\beta_c$~\cite{coddington}.}  
\label{fig2}
\end{figure}

In the presence of topological excitations local Monte Carlo simulations
may become inefficient at exploring the whole configuration space.
An example is provided by compact U(1) lattice gauge theory. In the Wilson
formulation, the action of the model is given by
\begin{eqnarray}
\label{action}
S = \beta \sum_{i, \mu < \nu} \left( 1 - cos (\theta_{\mu}(i) +
\theta_{\nu}(i + \hat{\mu}) - \theta_{\mu}(i + \hat{\nu})
- \theta_{\nu}(i)) \right) \ ,
\end{eqnarray}
where $\theta_{\mu}(i)$ is an angular variable defined on the link
stemming from the point $i$ in the direction $\mu$ and $\beta$ is
related to the bare coupling $g$ and to the lattice spacing $a$
by $\beta = 1/ag^2$. We shall consider the model in three dimensions.\\
In this model duality transformations can be performed, which amounts to
Fourier-transforming the Boltzmann weight~\cite{banks}. 
The dual theory is formulated in terms of a gas of Dirac magnetic monopoles.
By using the dual description, it can be proven that at zero temperature the
system permanently confines static charges for any value of the
coupling~\cite{polyakov,mack}. Using the same techniques, it can be proven
that at finite temperature permanent confinement is lost at some critical
value of the temperature $T_c$, above which the system is in a deconfined
phase~\cite{parga}. The phase transition is expected to be of the
Berezinskii-Kosterlitz-Thouless type. A numerical study of the phase
transition can be found in~\cite{coddington}.\\
In the direct formulation~(\ref{action}) magnetic monopoles appear as
topological excitations located at the centre of cubes and are detected as
sources of magnetic flux~\cite{degrand}. Flux conservation
is provided by Dirac strings ending at monopole locations.
The theory being compact, Dirac strings are undetectable.
Because of flux conservation, Dirac strings must either be attached to a
monopole or form closed loops. On a finite periodic lattice closed Dirac 
strings with non-trivial topology exist, which divide the lattice
into sectors. These are labelled by integers
representing the total wrapping number of closed strings in each
direction. An equivalent description of the topological sectors
can be performed in terms of fluxes~\cite{deforcrand}. In the presence
of an external flux $\phi$ across one plane the partition function is given by
\begin{eqnarray}
\label{zphi}
Z[\phi] =
\int_{\{\phi\}} \left( \prod_{i,\mu}{D} \theta_{\mu}(i) \right) \ 
e^{- S} \ ,
\end{eqnarray}
where $\{\phi\}$ indicates the set of all configurations carrying
flux $\phi$. Again because of compactness, external fluxes differing by
$2 \pi n$ are indistinguishable. The integer $n$ counts the
total number of closed Dirac strings piercing the lattice in the given
direction.\\
In numerical simulations of the model traditional algorithms prove to be
inefficient in exploring ergodicly the different topological
sectors~\cite{damgaard}. In this work we present a
cluster-like algorithm which is related to a recently introduced algorithm
for simulating bosonic systems \cite{alet1}. In our algorithm a proposed
update consists of a flux of randomly selected magnitude going through
plaquettes pierced by a
closed path in the dual lattice (we call that path a "worm"). The likelihood
of this update strategy being more efficient than local ones is based on the
fact that it should be more effective at forcing different units of flux into
the system. The basic steps of our algorithm can be summarised as follows:
\begin{enumerate}
\item a number $\varphi \in ]-\pi,\pi]$ is drawn with uniform probability;
\item a site $x_i = x_0$ of the dual lattice is randomly selected;
\item a priori probabilities to increase (respectively decrease) by
$\varphi$ the flux of each plaquette pierced by links stemming from
(resp. ending in) $x_0$ are computed;
\item a direction $\hat{\imath}$ is selected according to the probabilities
  computed in the previous step and its flux is updated accordingly;
\item steps 2 to 4 are iterated from the new point $x_n = x_{n-1} +
\hat{\imath}$ until $x_n$ is equal to the initial point $x_i$;
\item the new configuration is accepted with a probability $P_{\rm accept}$,
  related to the local configuration around $x_i$ in the starting and final
  configurations (see \cite{alet1} for more details). In most cases
  the acceptance rate is around 0.95-0.98.
\end{enumerate}  
The proof of detailed balance is analogous to the one given in
\cite{alet1}. An advantage of the ``worm'' algorithm over a local one is that
it changes the flux trough the system by generating worms
with non-trivial winding number, which should guarantee better ergodicity.
Moreover, our algorithm allows to determine $Z[\phi]$ (which is
proportional to the probability of having a flux $\phi$) for different
values of the flux.\\
As an application of the ``worm'' algorithm, we have studied $Z[\phi]$
across the deconfinement phase transition at finite temperature.
The expectation is that while in the confined phase
$Z[\phi]$ is independent of $\phi$, in the deconfined
phase higher fluxes through temporal planes are energetically
disfavoured. In fig.~\ref{fig1} we plot $Z[\phi]$ (normalised to 1) in
the confined and the deconfined phase. While in the former case ${Z}$ is 
independent of the flux, in the latter a clear peak at $\phi = 0$ emerges.
An interesting quantity is $Z[\pi]/Z[0]$, which is
one in the confined phase and goes to zero in the
deconfined phase. This ratio is plotted in fig.~\ref{fig2} as a function of
$\beta$: its sudden drop in the critical region is a clear signal of
deconfinement. A finite size study of $Z[\pi]/Z[0]$ is currently in progress.\\
In the critical region, our algorithm proves to be more efficient than
a local update at decorrelating topological sectors. For instance,
on a $32^2 \times 6$ lattice it gives shorter autocorrelation times
by a factor of 2-3. This factor is likely to increase on larger
lattices. This result might extend also to other observables like
the total number of monopoles, for which on a $64^2 \times 6$ lattice
we find $\tau_{worm} \simeq 2.0 \tau_{local}$ for $\beta \simeq \beta_c$.\\
~\\
In conclusion, we have presented a new algorithm for simulating compact U(1)
in 3d. Compared to local updates, our algorithm has the advantage of providing
better ergodicity across topological sectors and a clean way for measuring
ratios of partition functions. We have preliminary indication that close
to the deconfinement phase transition
our algorithm has shorter autocorrelation times than
local updates. The autocorrelation time of the ``worm'' algorithm can be
furtherly reduced by a more appropriate choice of the distribution of the
proposed flux or by reducing the backtracking of the worm,
using the concept of "directed worms"~\cite{syljua,alet2}.\\

\section*{Acknowledgments}
We thanks Ph. de Forcrand for discussions. While this paper was being
submitted, we were informed by T. Neuhaus that he has devised a
similar algorithm to ours, arriving to similar conclusions.
\bibliography{clusteru1}
\end{document}